\documentclass[final,5p,times,twocolumn]{elsarticle}
\usepackage[usenames,dvipsnames]{color}
\usepackage{graphicx}
\usepackage[utf8]{inputenc}
\usepackage[T1]{fontenc}
\graphicspath{{./img/}{./figs/}}

\journal{Physics Letters A}

\begin{document}
\begin{frontmatter}
\title{Currents in  metallic rings with quantum dot}
\author{Lukasz Machura\corref{} }
\author{Jerzy {\L}uczka \corref{cor1}}
\ead{jerzy.luczka@us.edu.pl}
\cortext[cor1]{corresponding author: tel. +48 32 359 1173; fax. +48 32 258 3653}
\address{Institute of Physics, University of Silesia, Katowice, Poland\\
Silesian Center for Education and Interdisciplinary Research, University of Silesia, 41-500 Chorz{\'o}w, Poland}

\begin{abstract}
Currents in a metallic ring with a quantum dot  are 
studied in the framework of a Langevin equation for a magnetic flux passing through the ring.  Two scenarios are considered: one in which thermal fluctuations of the dissipative part of the current are modelled by classical Johnson-Nyquist noise  and one in which quantum character of thermal fluctuations is taken into account in  terms of  a quantum Smoluchowski equation. The impact of the amplitude and phase of the transmission coefficient of the electron through a quantum dot on  current characteristics is analyzed. 
In  tailored parameter regimes, both scenarios can exhibit the transition from para-- to diamagnetic 
response of the ring current versus external magnetic flux.  
\end{abstract}


\end{frontmatter}

%
\section{Introduction}\label{intro}
In the early 90's after the successful reduction of the signal-to-noise ratio
the three groups conducted pioneering experiments with the mesoscopic metallic rings.
The careful measurements of Cooper \cite{LevDol1990}, Gold \cite{ChaWeb1991},
and Gallium-Aluminum-Arsenide/Gallium-Arsenide \cite{MaiCha1993} normal rings has 
shown the evidence of the existence of the persistent equilibrium currents 
flowing in the \emph{small metallic pieces of the rotational symmetry} 
reaffirming an old idea of Friedrich Hund \cite{Hun1938}. This very idea
concerns the charge transport in normal metallic ring. From the Ohm law we
can expect that from the macroscopic point of view such current will die out
within the relaxation time for a given material, which for metal is known to be 
rather short and of the order of $10^{-14}s$. However, for sufficiently small
circumferences the macroscopic description is no longer valid and ring reaches the
region where both macro-- and micro--world meet making the requirement for the 
mesoscopic description \cite{Imry1997} of the dynamics. In low enough temperature
the effects of quantum coherence of electrons appear. Under the right circumstances 
some electrons in the ring are able to preserve its coherence which in turn results
in a persistent (dissipationless) equilibrium current induced by the static magnetic 
field. In 1965 Bloch \cite{Blo1965} and five years later Kulik \cite{Kul1970} 
confirmed Hund's theory using the quantum-mechanical description.
The real interest in the topic of the persistent currents in normal rings arose 
after 1983 paper by Büttiker et. al \cite{ButImr1983} where the existence of 
the persistent currents was shown also in the presence of the elastic dispersions.

First measurements of currents in the diffusive regime \cite{ChaWeb1991} have
shown rather strong disagreement (10--200 times larger currents amplitudes) with the 
theoretically anticipated values. Later attempts reduced this dissimilarity 
to a factor of around 2-3 \cite{JarMoh2001}. Experiments with semiconducting
materials in the close to ballistic regime usually agreed with the theory 
\cite{MaiCha1993,RabSam2001}. Only recently the scanning SQUID technique was 
used to record not only the response signal of the rings itself but also from 
the background. This method gave the possibility of the high precision measurements 
of the current 
flowing in 33 different separate Gold rings \cite{BluKos2009} and finally confirm 
qualitatively as well as quantitatively all aspect of the existing theoretical
descriptions  
\cite{Imr2009}. The alternative method was used to measure the currents in the
Aluminum rings which were deposited on a cantilever \cite{BleSha2009}. 
A torque magnetometer whose vibration frequency can be precisely monitored was 
used as a detector. The measurements was performed with the
several different cantilevers decorated with a single aluminum ring or arrays of 
hundreds or thousands of identical Aluminum rings. The analysis of the different 
magnetic susceptibilities seen in \cite{BluKos2009,BleSha2009} based on the two--fluid model was addressed in \cite{DajLuc2003,MacRog2010,RogMac2010}. 

\begin{figure}[htbp]
\centering
\includegraphics[width=0.7\linewidth]{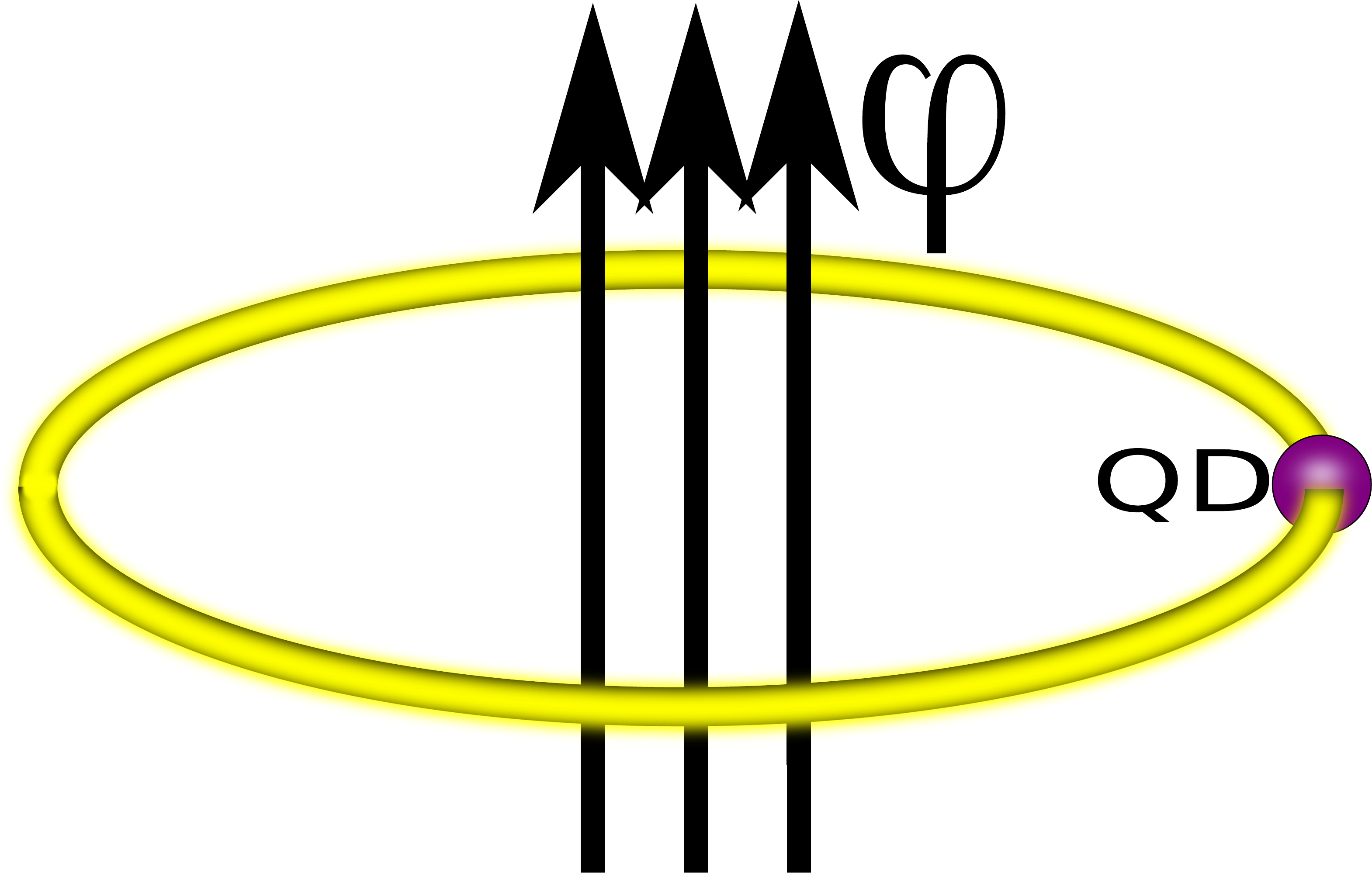}
\caption{Schematic picture of the mesoscopic non-superconducting  ring with a quantum dot QD. 
The ring is threaded by a magnetic flux  $\phi$. }
\label{fig1}
\end{figure}

In this work we present the analysis of electrical currents in the 
mesoscopic metallic (non-superconducting)  ring with the quantum dot. The experiment with the measurements 
of the phase of the transmission coefficient through a quantum dot in the Coulomb regime
was performed in 1995 \cite{YacHei1995}. Many different aspect of the persistent currents 
in the same scenario was studied rather intensively over the last two decades
\cite{LevBut1995,HacWei1996,WuGu1998,AffSim2001,BaoZhe2005}.
Similar schemes with the mesoscopic ring coupled to the quantum dot 
\cite{EckJoh2001,ChoKan2001,DinDon2003} or the quantum ring surrounding the quantum dot --
a dot-ring nanostructure (DRN) \cite{SomBie2011,SanSom2011,ZipKur2012} was also addressed. 
Here we follow the model proposed by Moskalets \cite{Mos1997} for a mesoscopic ring 
containing a potential barrier with a resonant level. 

The work is organized in the following way: In Sec. 2, the model is described and  the Langevin equation for the magnetic flux is presented both in the classical and quantum Smoluchowski regimes. Discussion of the results is presented in Sec. 3. 
 In Sec. \ref{subsec21}, the
stationary probability distribution  of the magnetic flux is analyzed. In Sec. \ref{subsec22},  the impact of parameters of the quantum dot on  average stationary currents is  studied and regimes of paramagnetic and diamagnetic response are worked out.  Sec. 4 contains  summary and conclusions.

\section{Flux dynamics of mesoscopic metalic rings with a quantum dot }

We consider a mesoscopic metallic ring in  an external magnetic field $B_e$
applied perpendicular to the plane of the ring.  
At zero temperature the ring can display a persistent current $I_P$ when  the size of the ring  is reduced to the scale of the electron quantum phase coherence length and the thermal length. At non-zero temperature $T>0$, a part of  electrons loses  phase coherence due to thermal fluctuations and
this part of  electrons contributes to a dissipative Ohmic current
$I_R$ associated with the resistance $R$ of the  metallic ring.
 The total  magnetic flux $\phi$ piercing the ring is a sum of the external flux $\phi_e\propto  B_e$  and the flux due to the flow of the current $I$, namely, 
\begin{equation}
\label{flux}
\phi = \phi_e + LI.
\end{equation}
Here,  $L$ stands for the self-inductance of the ring. The current $I$ is a sum of the persistent and dissipative currents,  
\begin{equation}
\label{I}
I=I_P + I_R. 
\end{equation}
Now, following Ref. \cite{Mos1997}, we assume that the ring contains a
potential barrier with a resonant level (a quantum dot). The expression for the persistent current $I_P=I_P(\phi)$ in such a system takes the form \cite{Mos1997}
\begin{eqnarray}
\label{IP}
I_P  &=& I_0\,   G(\phi/\phi_0) \sum_{n=1}^\infty  A_n (T/T^*)\cos[n(k_F l + \bar\delta_F)] \nonumber\\
&\times& \sin\{n \arccos[t_F \cos(2\pi \phi/\phi_0)]\}.   
\end{eqnarray}
The flux quantum $\phi_0=h/e$ is the ratio of the Planck constant $h$ and  the charge $e$ of the electron,  $I_0$ is the maximal persistent current at zero temperature for the ring without  the quantum dot and  
\begin{equation}
\label{G}
G(\phi/\phi_0) =  
\frac{t_F \sin(2\pi \phi/\phi_0)}{\sqrt{1 - t^2_F \cos^2(2\pi \phi/\phi_0)}} 
\end{equation}
modifies the maximal current  due to the quantum dot. Here, $t_F$ and $\bar\delta_F$  are the amplitude and phase of the transmission
coefficient  $T_k = t_k \exp[i\delta_k ]$ through a quantum dot for an electron of  the Fermi energy. 
 The  amplitudes $A_n(T/T^*)$ are determined by the relation
\begin{equation}
\label{A1}
A_n(T/T^*) =  \frac{T/T^*}{\sinh(n T/T^*)}, 
\end{equation}
where 
$T^*$ is the characteristic temperature which measures the level spacing at the Fermi surface. The magnetic flux $\phi$ is quantized with the flux quantum $\phi_0 = h/e$ being the ratio of the Planck constant $h$ and the electron charge $e$. Moreover,  $k_F$ is the Fermi momentum and $l$ is the circumference of the ring. 
For $t_F = 1$ and $\bar\delta_F = 0$ this expression  reduces to a current for a  pure metalic ring \cite{cheng}.

According to Ohm's law and Lenz's rule,  the dissipative current $I_R = I_R (\phi)$
assumes the form
\begin{equation}
\label{IR}
I_R = -\frac{1}{R} \frac{d \phi}{d t} + \sqrt{\frac{2 k_B T}{R}} \Gamma (t), 
\end{equation}
where $k_B$ is the Boltzmann constant. It means that we include  the effect of a nonzero temperature $T > 0$ by adding Johnson-Nyquist noise $\Gamma(t)$ which represents thermal fluctuations. They  are modeled by 
  $\delta$--correlated Gaussian white noise
of zero mean and unit intensity, 
\begin{equation}
    \label{Gama}
       \langle \Gamma(t) \rangle = 0, \quad \langle \Gamma(t)\Gamma(s) \rangle = \delta(t-s).
\end{equation}
Inserting  Eqs. (\ref{IR}) and (\ref{IP}) to the relation (1) yields
\begin{equation}
\label{Lan1}
 \frac{1}{R} \frac{d \phi}{d t}  = - \frac{1}{L}(\phi-\phi_e)  + I_P(\phi) + \sqrt{\frac{2 k_B T}{R}} \Gamma (t), 
\end{equation}
 We note that this equation is a Langevin equation for the magnetic flux $\phi =\phi(t)$. Indeed, it has the same form as a Langevin equation for an overdamped motion of a classical Brownian particle subject to the force $F = F(\phi)$ which reads  
\begin{equation}
\label{force}
 F(\phi) = -\frac{1}{L}(\phi-\phi_e)  + I_P(\phi)
\end{equation}
and the noise intensity strength $D=k_BT/R$ is in accordance with the classical 
fluctuation-dissipation theorem \cite{kubo66,zwan}. 
   Therefore we can apply the well-known  mathematical and numerical methods for  analysis of Eq. (\ref{Lan1}). First, we 
  transform it to the  dimensionless form (see \cite{DajRog2007,DajMac2007} for details)
\begin{equation}
\label{class}
\frac{dx}{ds} = - \frac{dV(x)}{dx} + \sqrt{2D_0}\;\xi (s).
\end{equation}
where $x = \phi/\phi_0$ is the dimensionless magnetic flux. The new time $ s= t/\tau_0$ , where the
characteristic time $\tau_0 = L/R$.  The thermal  noise intensity 
$D_0 =k_BT/(\phi_0^2/L) = (E_{T^*}/E_{\phi}) T_0 = k_0 T_0$, where the dimensionless temperature  $T_0 = T/T^*$, 
$E_{T^*} = k_B T^*/2$ is energy of thermal fluctuations at  the characteristic temperature $T^*$,  $E_{\phi} = \phi_0^2 / 2 L$ is the elementary magnetic energy 
and $k_0=E_{T^*}/E_{\phi}$ rescales intensity of thermal noise.   Rescaled  Gaussian white noise $\xi(s)$ has 
exactly the same statistical properties as the dimensional version $\Gamma(t)$.
The rescaled potential $V(x)$ takes the  form
\begin{equation}
\label{V}
V(x) = \frac{1}{2}(x - x_e)^2 + \alpha W(x).
\end{equation}
The rescaled external magnetic flux is denoted by $x_e = \phi_e/\phi_0$ and the nonlinearity parameter  $\alpha = L I_0 / 2 \pi \phi_0$. The potential  consists of the harmonic part $(x-x_e)^2/2$ and the periodic part 
\begin{eqnarray}
\label{W}
W(x) &=&  \sum_{n=1}^\infty \frac{A_n(T_0)}{n} \cos(n\delta_F) \nonumber\\
 &\times & \cos \{n \arccos[t_F \cos(2\pi x)]\},  
\end{eqnarray}
where $\delta_F = k_F l + \bar\delta_F$ is a shifted phase. 

The Fokker--Planck equation corresponding to the Langevin equation (\ref{class}) 
has the form  \cite{gard}
\begin{equation}\label{FP}
\frac{\partial}{\partial t} P(x, t) = \frac{\partial}{\partial x} \left[
\frac{dV(x)}{dx} P(x, t) \right] + D_0 \frac{\partial^2}{\partial x^2} P(x, t), 
\end{equation}
where $P(x, t)$ is a probability density of the process determined by Eq. (\ref{class}). From this equation, all statistical properties of the magnetic flux can be obtained. In particular, its statistical moments $\langle x^k(t) \rangle$ are determined by the expression  
\begin{eqnarray} 
\label{moments}
\langle x^k(t) \rangle = \int_{-\infty}^{\infty} x^k \; P(x, t) dx, 
\quad k=1,2,3, ...
\end{eqnarray}
For experimentalists, more interesting is the electrical  current flowing in the ring.  
 From Eq. (\ref{flux}) it follows that at any time the total current reads
\begin{equation}\label{I(t)}
 I(t) = \frac{1}{L} (\phi(t)-\phi_e)
\end{equation}    		
 and its  average value is given by the relation  
\begin{equation}\label{i(t)}
i(t) =  \langle x(t) \rangle - x_e, \quad i(t) 
= \frac{L}{\phi_0} \langle I(t)\rangle, 
\end{equation}    		
where  the dimensionless current $i(t)$ has been introduced. 
In the stationary state,  
\begin{eqnarray} 
\label{averagei}
i =  \langle x \rangle - x_e,\quad \langle x \rangle = \int_{-\infty}^{\infty} x \; P(x) dx,
\end{eqnarray}
where $P(x) = \lim_{t \to \infty} P(x, t)$ is a stationary probability density.  It  can easily be calculated from Eq. (\ref{FP}) for $\partial P(x, t) / \partial t = 0$ and zero
stationary probability current yielding the distribution
\begin{eqnarray}\label{Ps}
P(x) = \lim_{t \to \infty} P(x, t) = N_0 \exp\left[- \Psi_C (x)\right]  
\end{eqnarray}
and $N_0$ is the normalization constant. The generalized thermodynamic potential 
$\Psi_C(x) = V(x)/D_0$ depends on the external flux $x_e$ and the stationary
probability density is given by the Boltzmann distribution.  Eqs.  
(\ref{averagei})--(\ref{Ps}) form a closed set from which the non--linear function 
$i  = f(x_e)$ can be calculated determining the stationary current-flux characteristics.

\begin{figure*}[tbp]
\centering
\includegraphics[width=0.4\linewidth]{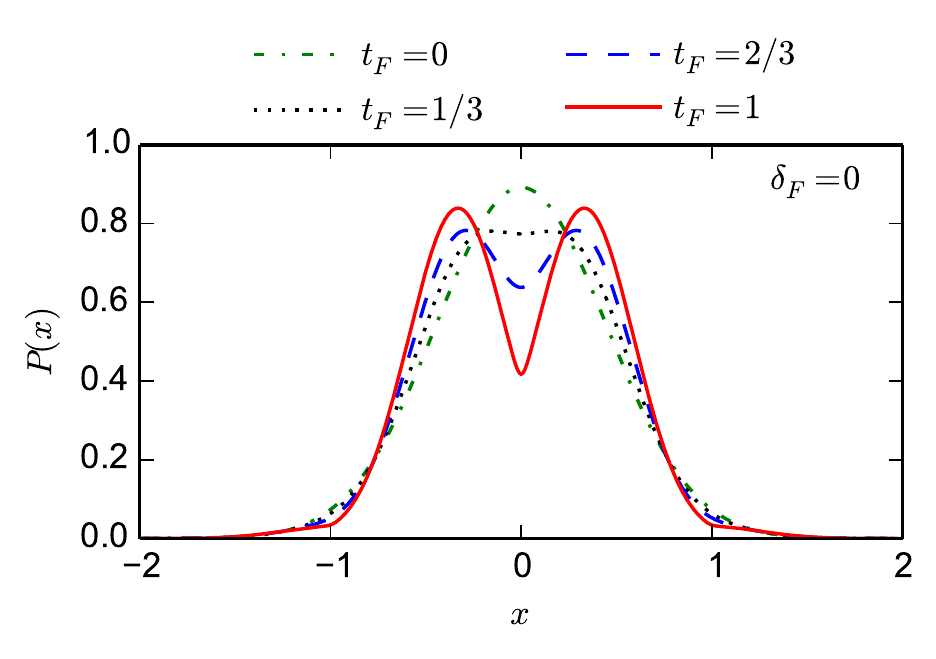}
\includegraphics[width=0.4\linewidth]{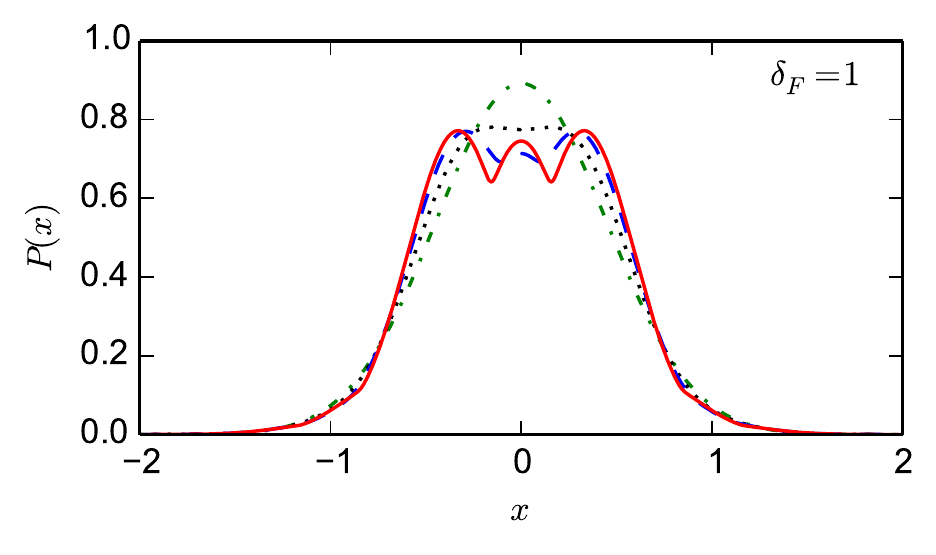}
\includegraphics[width=0.4\linewidth]{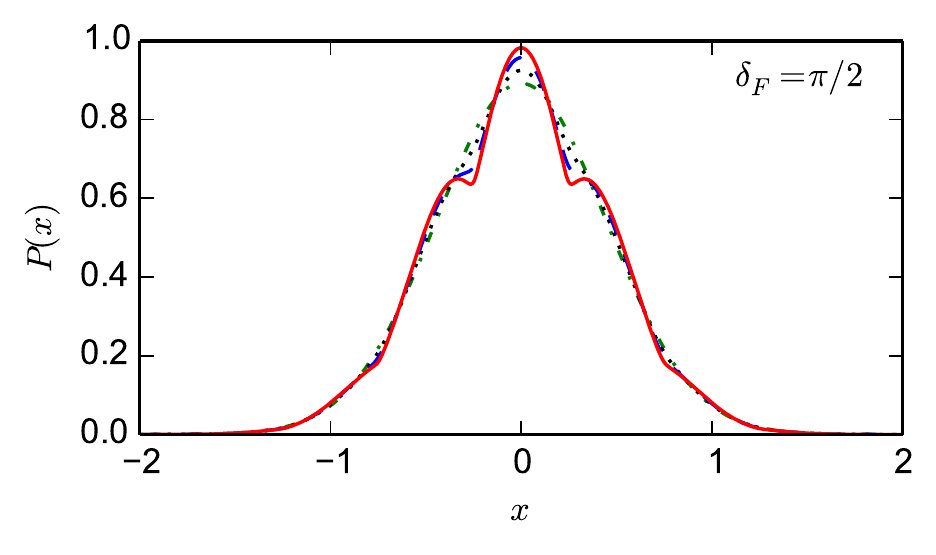}
\includegraphics[width=0.4\linewidth]{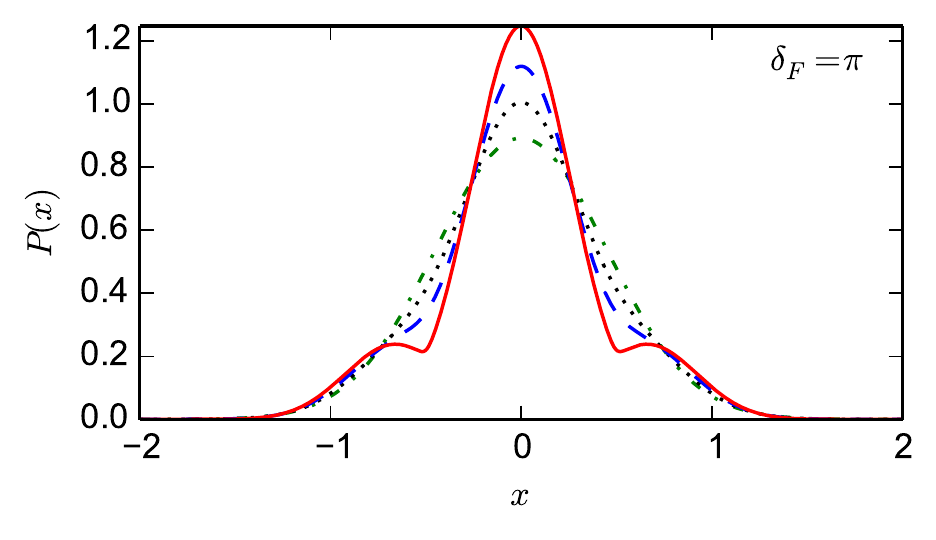}
\caption{The stationary probability distribution $P(x)$ of the dimensionless magnetic flux $x$ in  the classical
Smoluchowski regime for the external magnetic flux $x_e=0$ and four  values of the phase $\delta_F$ of the
transmission coefficient. 
In each  panel there are four curves which correspond to  different 
values of the amplitude of the transmission coefficient:  
$t_F = 0$ (green dashed--dotted), 1/3 (black dotted), 2/3 (blue dashed)
and 1 (red solid). Other parameters are:  $\alpha = 0.1$, $k_0 = 1$ and $T_0 = 0.2$.}
\label{cp}
\end{figure*}

\subsection{Quantum Smoluchowski limit}\label{subsec12}
Thermal fluctuations modeled as classical $\delta$-correlated white noise are adequate to describe many physical phenomena even in low temperatures. However, 
in some low temperature regimes,  quantum effects like tunnelling, 
quantum reflections and purely quantum fluctuations are playing an increasingly important  role and  quantum character of thermal fluctuations should be taken into account. How to do it is not a simple task and the problem in a general case  is still unsolved. In the so called quantum Smoluchowski limit, the leading quantum corrections are incorporated in the modified diffusion 
coefficient $D_0$ 
\cite{AnkPec2001,MacKos2004,RudLuc2005,
MacKos2006,MacKos2007,CofKal2008,CleCof2009,CofKal2009,AnkPec2010}. 
The modified diffusion coefficient takes the form \cite{MacKos2004}
\begin{equation}\label{Dlambda}
D_\lambda(x) = \frac{1}{\beta(1 - \lambda \beta V''(x))}, \quad \beta^{-1} =D_0.
\end{equation}
The prime denotes the differentiation with respect to $x$. The dimensionless quantum 
correction parameter 
\begin{equation}
\label{lambda}
\lambda = \lambda_0 \left[ \gamma + \Psi \left(1 + \frac{\epsilon}{T_0} \right) \right],
\end{equation}
where
\begin{equation}
\label{lamepsi}
\lambda_0 = \frac{\hbar R}{\pi \phi_0}, \quad 
\epsilon = \frac{\hbar}{2 \pi C R} \frac{1}{k_B T^*}.
\end{equation}
The psi function $\Psi$ is the digamma function (the logarithmic derivative of the 
gamma function). The $\gamma \approx 0.577$ is the Euler constant and $C$ is capacitance of the system 
related to the charging effects. The quantum correction parameter $\lambda$ is a difference between 
the quantum $\langle x^2 \rangle_q$ and classical $\langle x^2 \rangle_c$ second statistical moments  of the  magnetic flux (see Eq. (5) in Ref. \cite{AnkPec2001}),
\begin{equation}
\label{lambdafilo}
\lambda = \langle x^2 \rangle_q - \langle x^2 \rangle_c.
\end{equation}
The modification of the diffusion coefficient (\ref{Dlambda}) results in modification of the Langevin equation, namely, 
\begin{equation}
\label{quant}
\frac{dx}{ds} = - \frac{dV(x)}{dx} + \sqrt{2D_\lambda(x)}\;\xi (s)
\end{equation}
and should be interpreted in the Ito sense \cite{gard}. The corresponding Fokker-Planck equation has the form 
\begin{equation}\label{FPQ}
\frac{\partial}{\partial t} P(x, t) = \frac{\partial}{\partial x} \left[
\frac{dV(x)}{dx} P(x, t) \right] +  \frac{\partial^2}{\partial x^2} 
\left[ D_{\lambda}(x)P(x, t)\right].  
\end{equation}
The stationary solution of this equation reads 
\begin{equation}
\label{PQ}
P(x) = N_0 D^{-1}_\lambda(x) \exp[-\Psi_\lambda(x)],
\end{equation}
where  the generalized thermodynamic potential takes the form 
\begin{equation}
\label{gt}
\Psi_\lambda (x) = \beta V(x) - \frac{\lambda \beta^2}{2} [V'(x)]^2,
\end{equation}
We emphasize that the stationary distribution describes an  equilibrium state, but it is not a Gibbs state. Remember that the Gibbs state is correct in the limit of a weak coupling of the system with thermostat. The Smoluchowski limit corresponds to the strong coupling regime.

\begin{figure*}[tbp]
\centering
\includegraphics[width=0.4\linewidth]{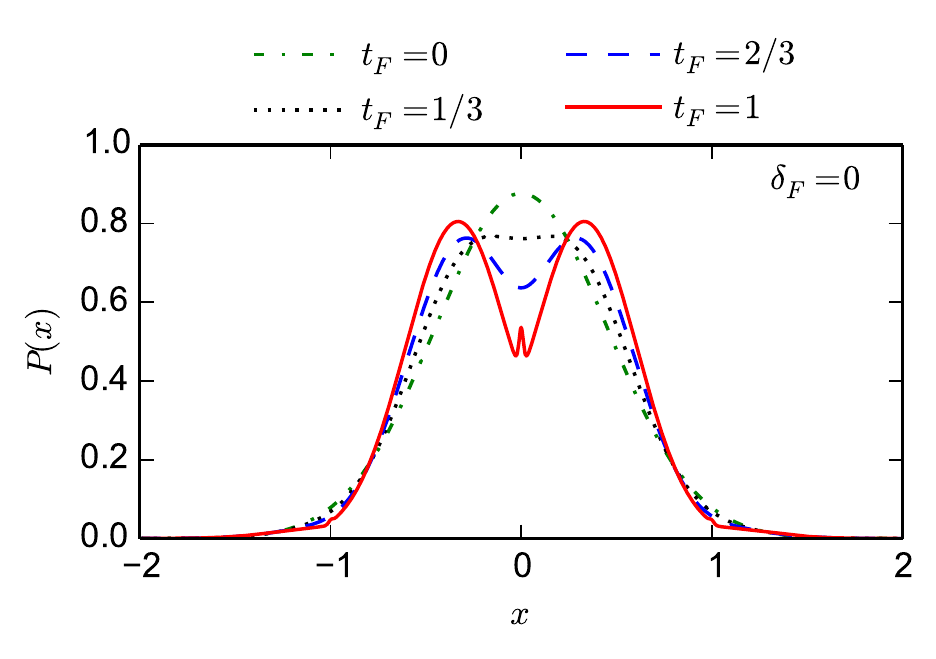}
\includegraphics[width=0.4\linewidth]{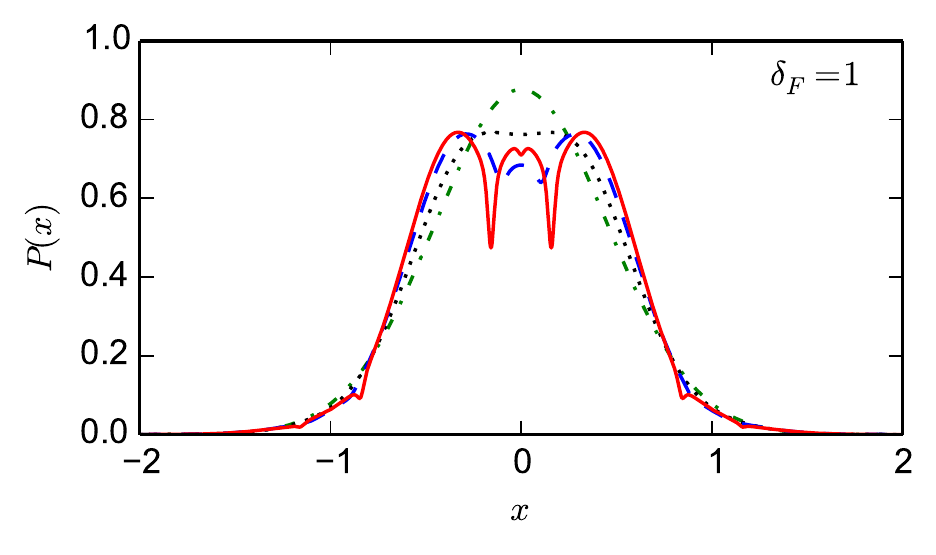}
\includegraphics[width=0.4\linewidth]{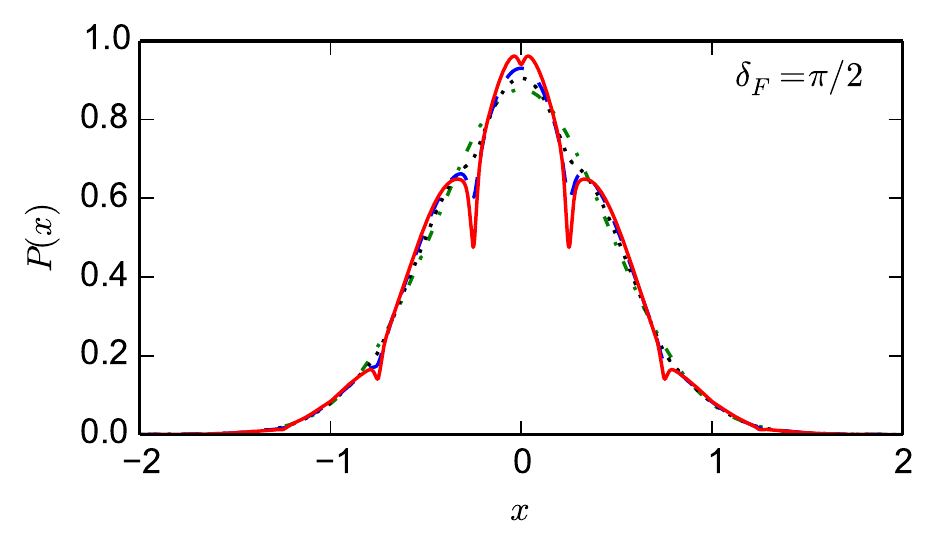}
\includegraphics[width=0.4\linewidth]{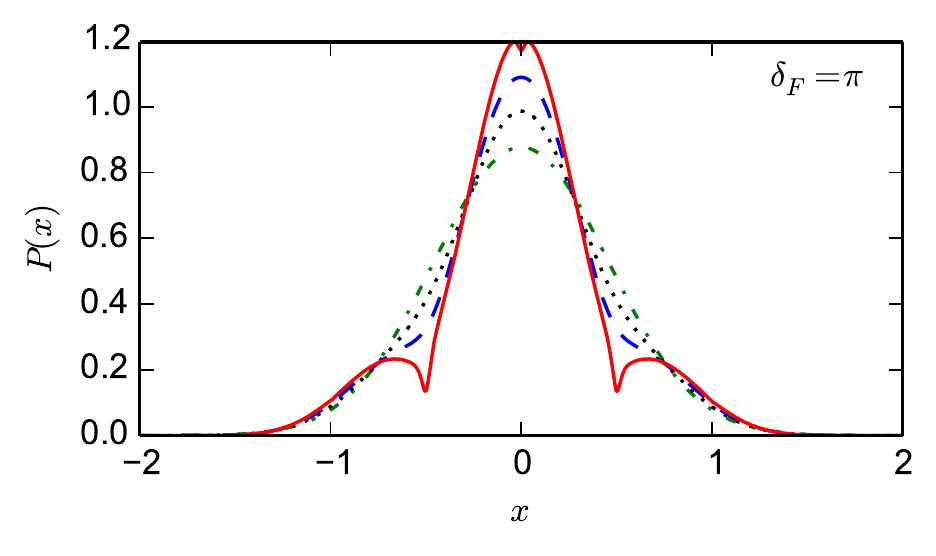}
\caption{The stationary probability distribution $P(x)$ in  the quantum
Smoluchowski regime  for $x_e=0$ and four  values of the phase $\delta_F$ of the
transmission coefficient. In each  panel there are four curves which correspond to   selected values of the amplitude of the transmission coefficient:  
$t_F = 0$ (green dashed--dotted), 1/3 (black dotted), 2/3 (blue dashed)
and 1 (red solid). The remaining parameters read $\alpha = 0.1$, $k_0 = 1$, $T_0 = 0.2$, $\lambda_0=0.001$ and
$\epsilon=100$.
}
\label{qp}
\end{figure*}

\section{Discussion of results} \label{sec2}

For the ring without a quantum dot, our model reproduces  experimental
data both for the diamagnetic and paramagnetic response in the vicinity of 
 zero magnetic field \cite{MacRog2010}. For the ring with a quantum dot, we have not found experimental data. Therefore,  our work could inspire experimentalists to design experiments and verify our
theoretical predictions revealed below:  the influence of the transmission
coefficient $t_F$ and the phase $\delta_F$ of the quantum dot on stationary current-flux characteristics.

The system has a 8-dimensional parameter space 
$\{x_e, T_0, k_0, \alpha, \lambda_0, \epsilon, t_F, \delta_F\}$. 
It would be difficult to carry out a comprehensive analysis and present current-flux characteristics for all possible sets of parameters.  Therefore, 
for  numerical calculations,  values of the  parameters 
$\alpha = 0.1$, $k_0 = 1$ and $T_0 = 0.2$ are kept fix. We include  quantum corrections which are  characterized by 2 parameters: $\lambda_0$ and $\epsilon$. Their physical meaning is explained in Refs. \cite{DajRog2007, DajMac2007}.  The quantum dot is also characterized by 2 parameters: $t_F$ and $\delta_F$ and their impact is displayed below. 
they will be fixed at the value of $\lambda_0 = 0.001$ and $\epsilon=100$.
The similar analysis  but for the pure metallic ring without the quantum dot
is  presented in our previous papers. The stationary solutions of the Fokker--Planck
equation was addressed in Ref. \cite{DajMac2007} and the current--flux characteristics
was investigated in Refs. \cite{MacRog2010,RogMac2010}.

\subsection{Stationary states}\label{subsec21}
The stationary solution of the Fokker--Planck equations (\ref{FP}) and (\ref{quant}) is given 
by the steady-state  probability distribution $ P(x) $ through the relations 
(\ref{Ps}) and (\ref{PQ}) without and with the quantum corrections, respectively. 
We consider the case $x_e=0$, i.e. when the external magnetic field is absent. 
In the case of classical thermal fluctuations, the Boltzmann distribution is  depicted in Fig. \ref{cp} for four different values of the phase of the
transmission coefficient  $\delta_F=0$ (top--left),
$1$ (bottom--left), $\pi/2$ (top--right), $\pi$ (bottom-right). 
For $x_e=0$, the probability distribution is symmetric with respect to the reflection $x \to -x$. Moreover, it is invariant under the  change of the phase
$P(x, \delta_F) = P(x, 2 \pi - \delta_F)$. Therefore below we consider the interval of the phase $\delta_F \in [0, \pi]$. 
All four panels present the distributions for four different
transmission coefficient $t_F = 0, 1/3, 2/3, 1$.

For  full transmission (i.e. $ t_F=1 $) the distribution
possesses two local maxima for low valued phases, which reflects 
the bistability of the generalized thermodynamic potential $ \Psi_C $. 
This, in turn, means that in the steady-state the  current can flow in two direction: clockwise or  counterclockwise (but the averaged current is zero!).   
For the phase $\delta_F \simeq 1$ and  full transmission 
three local maxima can be found, with the most probable 
aside the local maximum around $x=0$ (which denotes the zero current
state). 
The additional local extrema, which doesn't appear in the
$\delta_F \to 0$ case, indicate the possible multi--stability.  
This means that again the 
self--sustaining persistent currents can appear without the applied magnetic 
flux and are more probable than the zero current state.
For the moderate--to--high phases the local maximum of the
probability distribution at $x=0$ becomes
the most protruding among all others located at more distant values of
the flux $x$. It means that  self--sustaining
currents are difficult to induce. Moreover, the lifetimes of the
induced currents related to the remote from zero extrema are also expect to be 
relatively short \cite{DajMac2007}.

\begin{figure*}[tbp]
\centering
\includegraphics[width=0.4\linewidth]{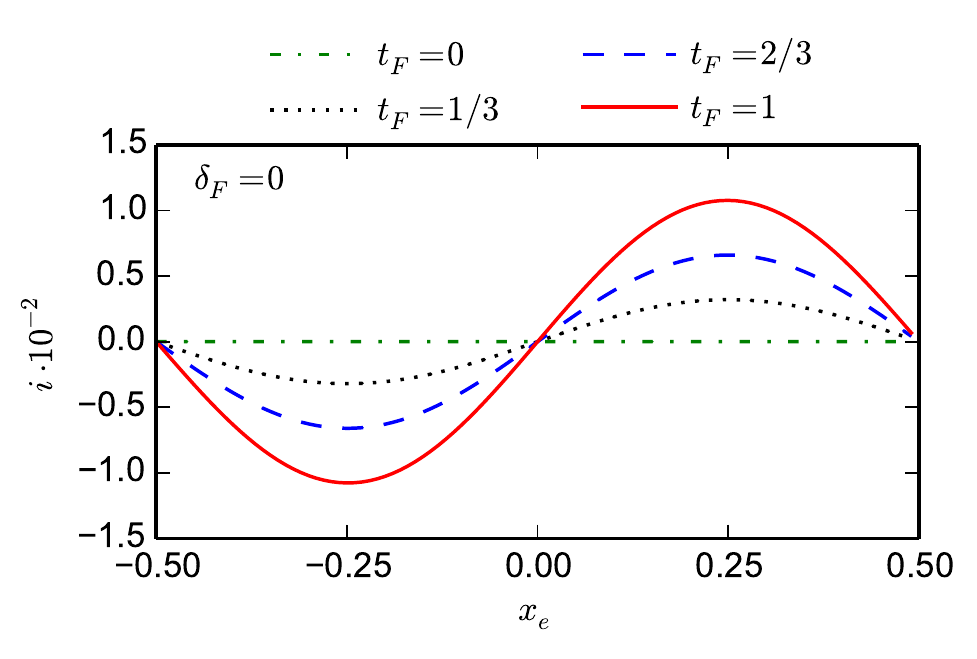}
\includegraphics[width=0.4\linewidth]{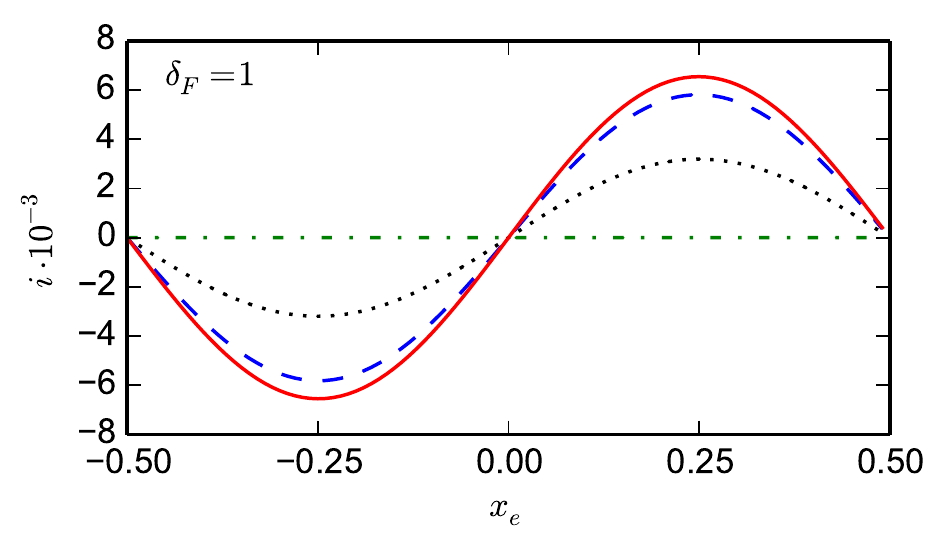}
\includegraphics[width=0.4\linewidth]{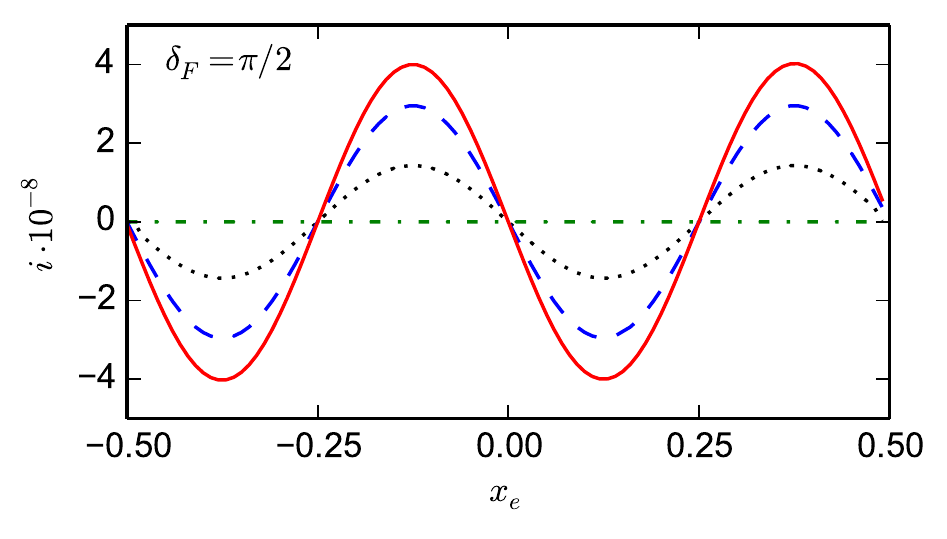}
\includegraphics[width=0.4\linewidth]{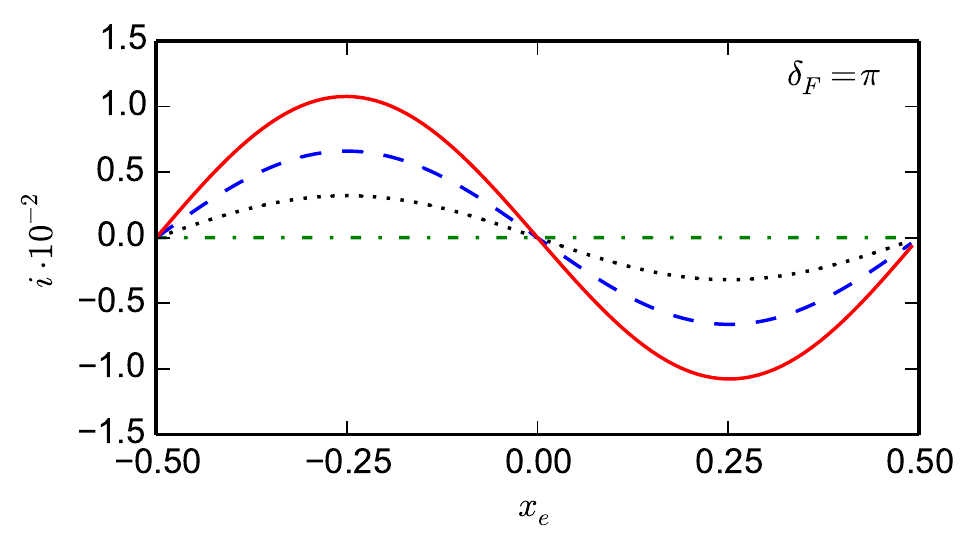}
\caption{The stationary averaged  current $i$ in the classical Smoluchowski  regime  
versus the external magnetic flux $x_e$ is presented for four different values of the phase $\delta_F$.  In each  panel there are four curves which correspond to   selected values of the amplitude of the transmission coefficient: 
$t_F = 0$ (green dashed--dotted), 1/3 (black dotted), 2/3 (blue dashed)
and 1 (red solid). Other system
parameters read $\alpha = 0.1$, $k_0 = 1$, $T_0 = 0.2$. }
\label{ci}
\end{figure*}

As already stated, in the quantum Smoluchowski limit, the stationary solution (\ref{PQ}) describes
the thermodynamic equilibrium. It is not, however, the quantum Gibbs state,
as we deal with the strong coupling to the environment. In this case,  the probability distribution depends explicitly on the coupling of the ring with thermostat via the resistance $R$ in the parameter $\lambda_0$ in Eq. (\ref{lamepsi}). 
The equilibrium stationary distribution with quantum corrections is depicted in Fig.  \ref{qp} for the same set of the parameters
as in the classical counterpart in Fig. \ref{cp}. For  the quantum 
corrections we set $ \lambda_0 = 0.001 $ and $ \epsilon = 100 $.
This means that the difference between the quantum and classical fluctuations 
of the dimensionless magnetic flux is  $\lambda = 0.0075 $. 
The  corrected distribution display somehow
magnified features seen in the corresponding classical cases: minima are deeper and maxima are more pronounced.

\begin{figure*}
\centering
\includegraphics[width=0.4\linewidth]{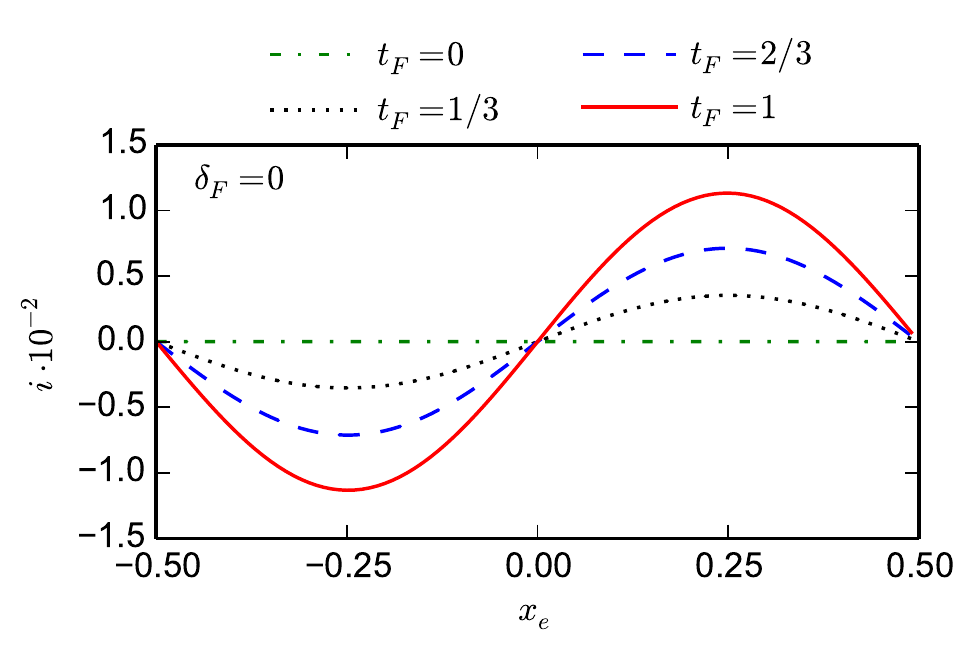}
\includegraphics[width=0.4\linewidth]{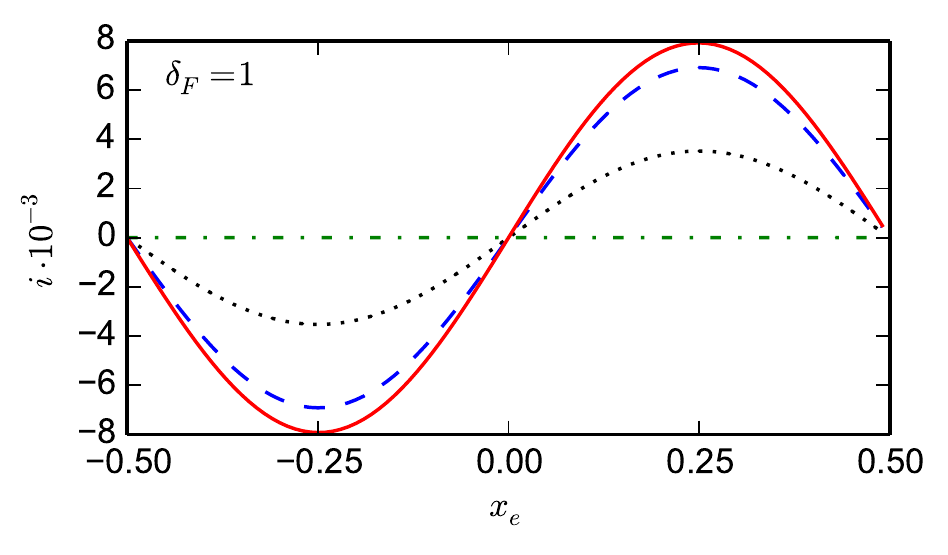}
\includegraphics[width=0.4\linewidth]{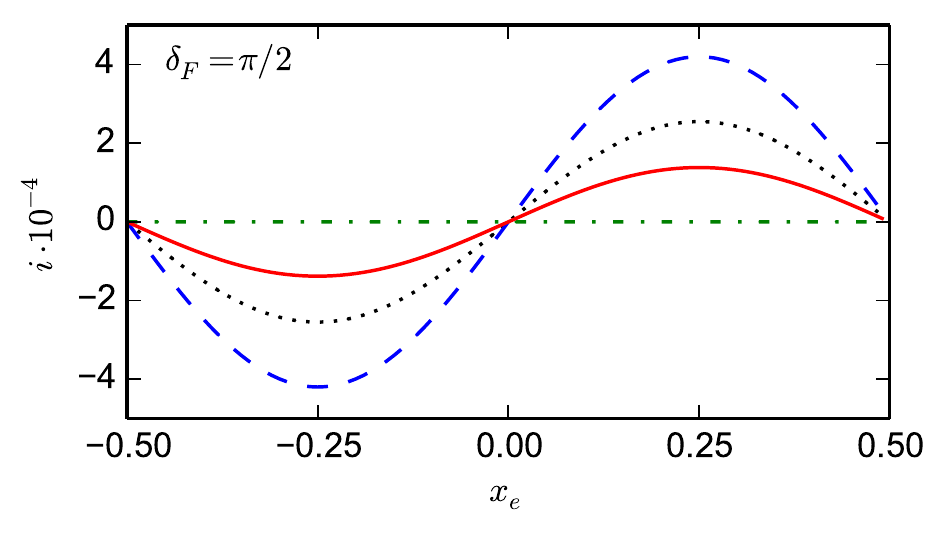}
\includegraphics[width=0.4\linewidth]{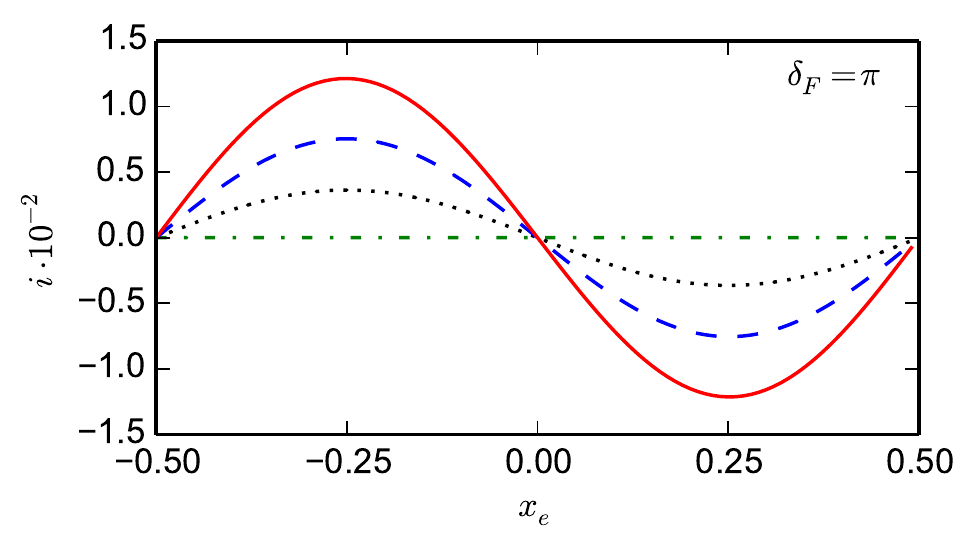}
\caption{The stationary averaged current $i$ in the quantum Smoluchowski regime 
versus the external magnetic flux $x_e$ is presented for 
four  values of the phase $\delta_F$ and  four  
values of the amplitude of the transmission coefficient:  
$t_F = 0$ (green dashed--dotted), 1/3 (black dotted), 2/3 (blue dashed)
and 1 (red solid). Other system
parameters read $\alpha = 0.1$, $k_0 = 1$, $T_0 = 0.2$, $\lambda_0=0.001$ and
$\epsilon=100$.}
\label{qi}
\end{figure*}

\subsection{Current--flux characteristics}\label{subsec22}

In previous papers \cite{MacRog2010,RogMac2010}, impact of quantumness of thermal fluctuations on the current-flux characteristics has been studied.  
In this section we will focus on influence of the quantum dot on such characteristics. For zero external magnetic flux, $x_e=0$, the averaged stationary current is zero. It follows from the properties of the stationary distribution: it is an even  function of $x$.  The non--zero magnetic flux  $x_e \neq 0$  breaks 
the $x$--reversal symmetry and the non-zero averaged current can emerge. 
In  Fig. \ref{ci} we depict  the response of the metallic ring 
to  the applied constant magnetic flux in  the classical Smoluchowski regime 
 (i.e. for $\lambda = 0$). It is worth to stress that the 
current characteristics for the amplitude $t_F = 1$ and the phase $\delta_F = 0$ of the transmission coefficient (top panel, red curve)
represent the situation with maximal current. In other words, it is the same as the ring   without quantum 
dot. The suppression of the generated signal which comes with the
reduction of the transmission amplitude seems to be the usual situation.
For $t_F = 0$ it is impossible to generate current in the ring. 
For  the phase $\delta_F \in (-\pi/2, \pi/2)$   the current response 
of the ring is paramagnetic for all non-zero amplitudes $t_F$. 
In turn, for $\delta_F \in (\pi/2, 3\pi/2)$ the response is diamagnetic. 
Let us note the doubled  period  for the particular case $\delta_F=\pi/2$. The analysis for slightly
lower or higher phases shows simple para- or diamagnetic single--periodic 
structure of current--flux characteristics, respectively.

\begin{figure}[t]
\centering
\includegraphics[width=1\linewidth]{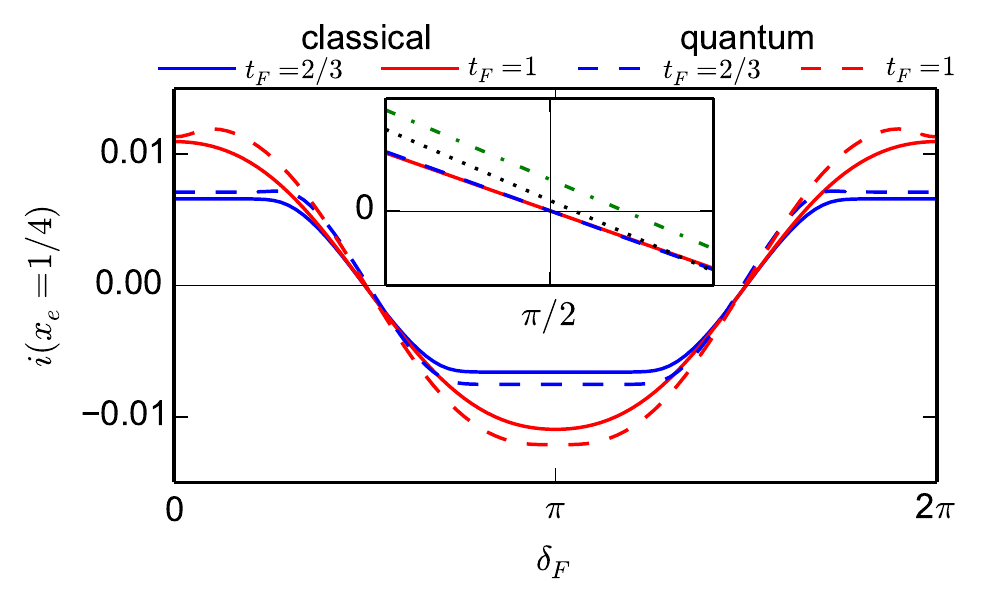}
\includegraphics[width=1\linewidth]{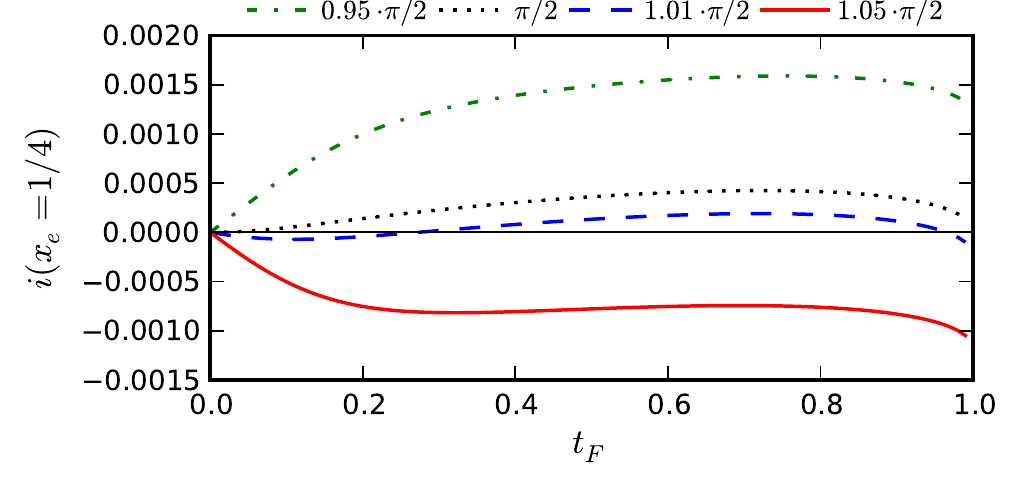}
\caption{The regimes of paramagnetic ($i>0$) and diamagnetic ($i<0$) response
in dependence of the phase $\delta_F$ and the amplitude $t_F$  of the transmission coefficient through a quantum dot for  an electron of the Fermi energy. The external magnetic flux is fixed at $x_e=1/4$. For the phase close to $\pi/2$ the current reversal is observed (upper panel). 
In bottom panel, the stationary averaged current in the quantum Smoluchowski regime is depicted as a function of  the amplitude $t_F$ and  in the vicinity of the phase $\delta_F=\pi/2$.  Four curves correspond to four values of the phase $\delta_F = 0.95 \pi/2, \pi/2, 1.01 \pi/2, 1.05 \pi/2$. 
Other system parameters read $\alpha = 0.1$, $k_0 = 1$, $T_0 = 0.2$, 
$\lambda_0=0.001$ and $\epsilon=100$.}
\label{qi2}
\end{figure}

We now  address the issue of whether, and to which extent,
the quantum nature of thermal fluctuations  can influence transport properties. 
We thus show impact of quantum corrections on the current  characteristics 
in Fig. \ref{qi} for the fixed  
quantumness  parameters $\lambda_0=0.001$ and $\epsilon = 100$. This figure is organized in exactly the same way as the previous one although the peculiarities are slightly
different. For instance we cannot conclude here, that the maximal possible current
amplitude is typically realized for $t_F = 1$. 
In the classical Smoluchowski regime, the case  $t_F=1$ is always the most
optimal.  With quantum  corrections, it is intriguing to note that  
around  $\delta_F=\pi/2$ 
the maximal amplitude of the transmission coefficient does not provide
maximal current. In fact the  current is weaker for $t_F=1$ than 
for $t_F=2/3$ or even when $t_F=1/3$, see Fig. 5. In fact one can observe something similar
to the transition from the paramagnetic to the diamagnetic state simply
by changing the phase around $\delta_F=\pi/2$. This is displayed in
Fig. \ref{qi2}. For the phases a little bit higher than $\pi/2$, like one 
identify the classical picture -- c.f. bottom panel on Fig. \ref{qi}.

As the next point of analysis we ask about   domains of parameters $\delta_F$ and $t_F$   where the current is paramagnetic and diamagnetic, see Fig. \ref{qi2}.   In  the  case of classical thermal fluctuations, the current is always of a paramagnetic type  in the interval $\delta_F \in (0, \pi/2)$ and  is always of a diamagnetic type  for $\delta_F \in (\pi/2, 3\pi/3)$. In the case of quantum  thermal fluctuations, it is not true: these intervals depend on the amplitude of the transmission coefficient. Nevertheless,  the current is paramagnetic in a large interval around $\delta_F =0$ and is diamagnetic in a large interval around $\delta_F = \pi$, and the transition point is in a small interval around $\delta_F =\pi/2$. The type of response 
is more robust to changes in the amplitude of the transmission coefficient and more sensitive to changes of the phase around the value $\pi/2$.

\section{Summary} 

This paper presents  the influence of the quantum dot on transport properties of mesoscopic non-superconducting rings. The theory is constructed in the framework  of the Langevin equation for  magnetic flux dynamics. We have considered the case when the system is driven by  classical thermal noise in the Smoluchowski regime. The so named quantum Smoluchowski regime has also been studied. The  stationary probability  distribution both in 'classical' and 'quantum' case is  depicted for   zero external magnetic flux. The current-flux characteristics are   analyzed in detail. The impact  of parameters characterizing the quantum dot on the current has been addressed in this work. The phase of the transmission coefficient plays the crucial role in type of the response. In the 'classical'  case,  its  crossover value  is fixed to $\delta_F = \pi/2$. Below this value, the current is paramagnetic while above this value the current is diamagnetic. For the 'quantum' case, the response threshold depends on other parameters of the system, nevertheless it is located close to the value $\pi/2$. Finally, we  would like to mention that recent progress in entirely novel experimental techniques makes the verification of our findings possible and we hope that our work will contribute  to the development of effective control methods of transport properties in mesoscopic systems. 


\section*{Acknowledgement}
The work supported in part by the NCN grant DEC-2013/09/B/ST3/01659.

\end{document}